\pdfoutput=1
\documentclass[11pt]{article}
\usepackage[pdftex]{graphicx}
\usepackage{caption} 
\usepackage{cite}
 
\begin{document}

\title{\sc Two-dimensional Ultrasound Imaging Technique based on Neural Network using Acoustic Simulation}
\author{Yoshiki Nagatani \thanks{Research and Development Department, Pixie Dust Technologies, Inc. 2-20-5, Kanda-Misakicho, Chiyoda-ku, \newline \hspace*{0.5cm} Tokyo 101-0061, Japan, yoshiki.nagatani@pixiedusttech.com} \and Shigeaki Okumura \thanks{MaRI Co., Ltd., Kyoto, 600-8815, Japan} \and Shuqiong Wu \thanks{The Institute of Scientific and Industrial Research, Osaka University, Osaka, 567-0047, Japan} \and Tetsuya Matsuda \thanks{Graduate School of Informatics, Kyoto University, Kyoto, 606-8501, Japan}}

\date{}
\maketitle


\begin{abstract}
\noindent The two-dimensional (2D) ultrasound imaging is widely used in several fields. In this study, the potential of two-dimensional ultrasound imaging technique based on machine learning is shown. A line transmitter was placed at one side and a receiver array was placed at the opposite side of the medium, respectively. The received signals at the receiver array was considered as a 2D image and it was converted to a 2D geometry of the target medium using convolutional neural network (CNN). The results of the numerical simulation studies which use up to 40000 simulation data showed the good agreement between the estimated and true geometry. 
\\
\\
\noindent {\bf Keywords:} FDTD simulation, acoustic imaging, CNN, cancellous bone.
\end{abstract}


\section{Introduction}

 In this study, we focus on the two-dimensional (2D) ultrasound imaging technique based on machine learning techniques. The 2D ultrasound imaging is used in several applications such as non-invasive imaging in the medical field and non-destructive evaluations for the constructions in the civil engineering field \cite{1,2}. However, due to the non-negligible artifact, the imaging techniques is not widely used for the measurement of the medium which has complex geometry, very different values of the acoustic impedance \cite{3,4} and wave speed, such as the cancellous bone which exists at the inner part of the bone and has the honey-comb-like structure.
 
The evaluation of the cancellous bone is now conducted by evaluating the ultrasound signals propagate along the cancellous bone. The device normally has a pair of a transmitter and a receiver. The evaluation of the amplitude and propagation speed of the ultrasound leads the estimation of the bone parameters such as Bone Volume Fraction (BV/TV), or porosity \cite{5,6}. In addition to the estimation of the BV/TV value, for example, the method of the wave separation by parameter fitting \cite{7,8,9}, the robust estimation of the instantaneous frequency \cite{10}, the precise investigation of attenuation slope \cite{11}, and a lot of other techniques were proposed to characterize the bone property by using the experimental measurements and also by the numerical simulations. However, the imaging of the medium is still difficult to achieve.

Recently, we applied conventional machine learning based techniques for the BV/TV estimation with simulation study \cite{12,13}. The estimation results showed the good correlation with the true BV/TV, which is the scalar value. The previous study shows the potential of the evaluation of complex medium with ultrasound signal. In other words, the result tells us that the ultrasound probably contains the precise information of the medium where the wave has been propagating. 

In this study, we applied machine learning techniques for the 2D imaging assuming that the waveforms contain rich information of the medium. The machine learning technique is expected to extract, convert, and retrieve the information from the reasonably related sources. 
Recently, CNN (Convolutional neural network) is proved considerably effective in image retrieval and feature extraction \cite{14,15}. For example, CNN-based features were applied to the classification and retrieval of food images \cite{14}. Similarly, in an iris recognition problem, CNN was used to extract robust features \cite{15}. Here, we know that ultrasound signal propagates along the medium based on wave equation. Therefore, the relationship between the ultrasound signal and the medium may be easily learned by CNN. If we provide enough and appropriate information to the CNN, we can inversely solve the wave equation and retrieve the 2D arrangement of the medium.

In this paper, as the pilot study, we show the potential of the machine learning based 2D imaging technique with numerical simulation study. To get good results from the trained CNN and evaluate the network, we need to prepare a certain number of training and test data. Here, the simplified and easily implementable setup is suitable for the purpose of this pilot study. We placed a single transmitter at the upper side of the medium and received with an array at the bottom of the medium. We did not need any preprocess for example a time-delay process that makes a focal point. We applied the received signals to the neural network (NN) directly and converted them to the 2D imaging result.


\section{FDTD Simulation Setting}

\begin{figure}[htb]
  \begin{center}
    \includegraphics[width=12.0cm]{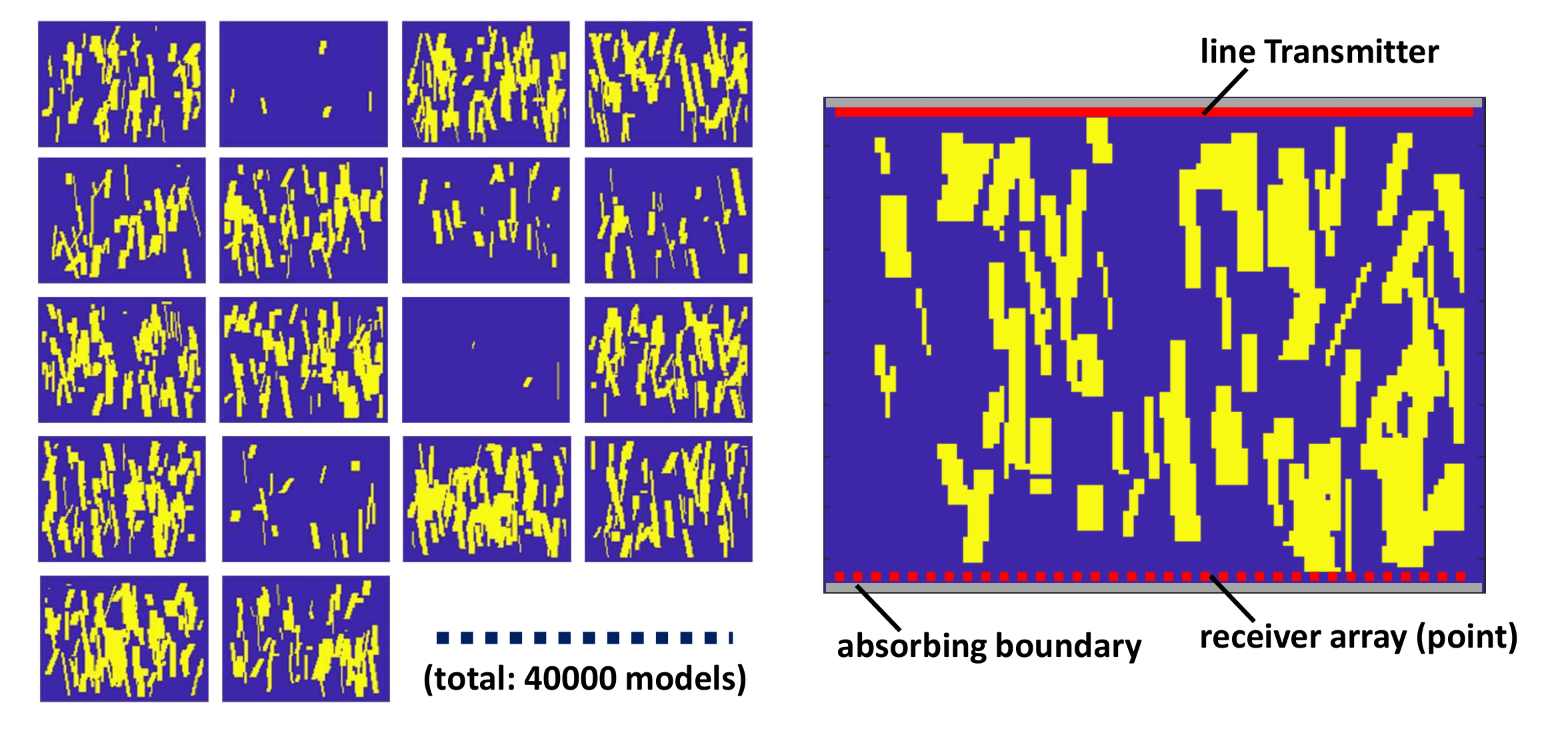}
    \captionsetup{margin=1.0cm,font={sc, footnotesize}}
    \caption{Schematic illustration of (a) simulated propagation media and (b) simulation setting. The line transmitter is placed at top of the medium and the receiver array is placed at bottom of the medium.}
    \label{fig:fig1}
  \end{center}
\end{figure}

For the simulation, 40000 2D geometries, whose size was $96 \times 128$ pixels, were prepared. To mimic the trabeculae of the cancellous bone, rod-like reflectors were located randomly in the field (length = $16 \pm 8$ px, thickness = $3 \pm 2$ px, angle = $0 \pm 10$ degrees). The space occupation ratio was between 0.0 and 0.5. The absorbing boundary condition was implemented at the two edges at the sides of the transmitter and receiver array of the model. The lateral edges were assumed as total reflecting walls. Figure 1 (a) shows examples of the geometry. Using the models, the wave propagation was calculated using the 2D acoustic finite-difference time-domain (FDTD) simulation \cite{16}. Here, shear wave and the absorption of the material were not considered. A single pulse of a sinusoidal wave was transmitted from a line source. Then, the propagated waves were received at the 64 points located on the opposite side of the transmitter as shown in Fig. \ref{fig:fig1} (b).

\section{Retrieval of 2D arrangement with CNN and training setting}

\begin{figure}[htb]
  \begin{center}
    \includegraphics[width=13.0cm]{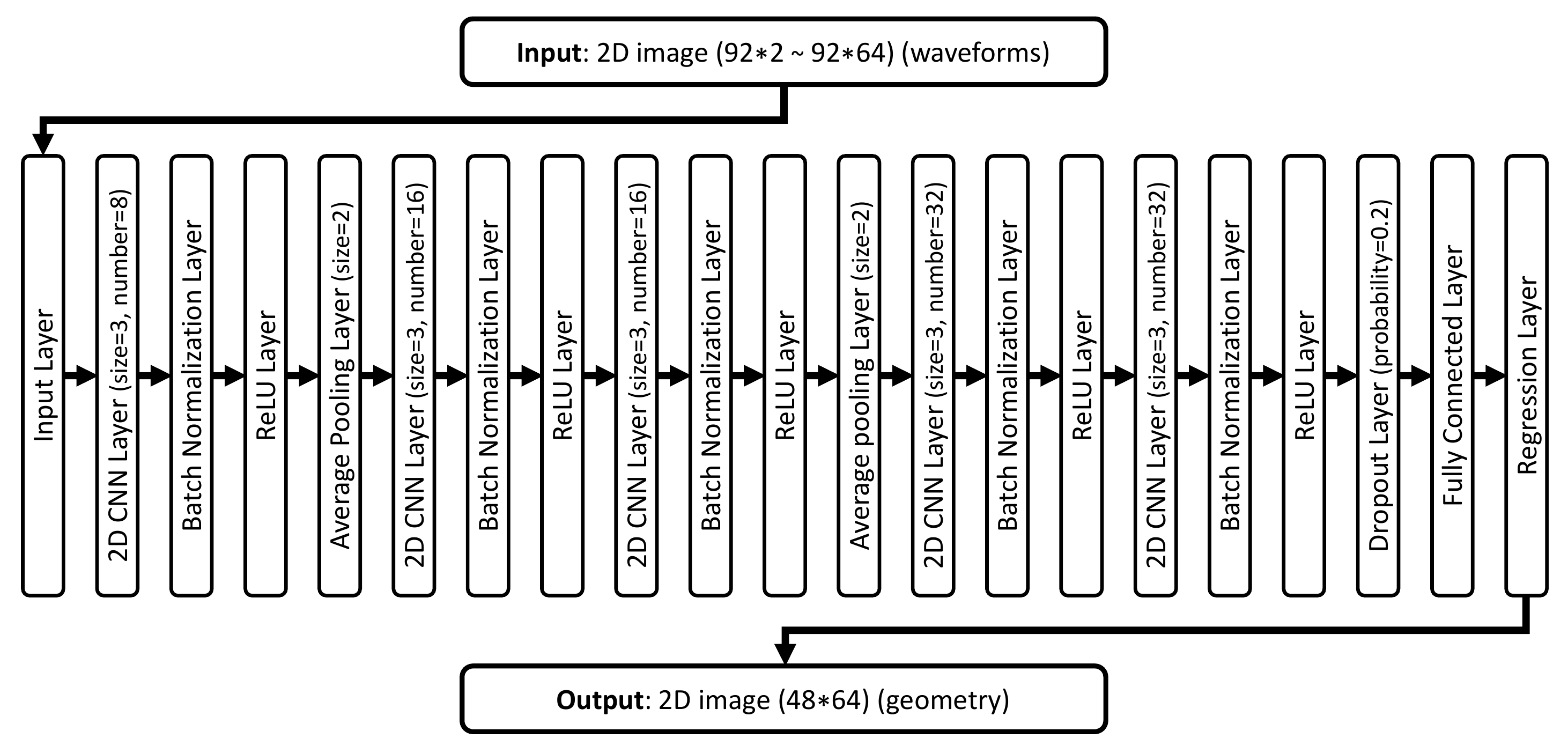}
    \captionsetup{margin=1.0cm,font={sc, footnotesize}}
    \caption{The structure of the neural network used in this study.}
    \label{fig:fig2}
  \end{center}
\end{figure}

The pairs of the waveforms and the geometry were used for training. In this study, the CNN was used to find the relationship between the geometry and waveforms. The received waveforms at the receiver array were used as the input data, and the shrank images of the original geometry ($48 \times 64$ pixels) were used as the output data. The 95\% of the derived data were used for training and validation, and the remaining 5\% were used for testing. The structure of the CNN we used in this study is shown in Fig. \ref{fig:fig2}. It contains one input layer, 5 convolutional layers, 2 average pooling layers, 5 rectified linear unit (ReLU) layers, and 1 dropout layer.

\section{Result and discussion}

\begin{figure}[htb]
  \begin{center}
    \includegraphics[width=11.0cm]{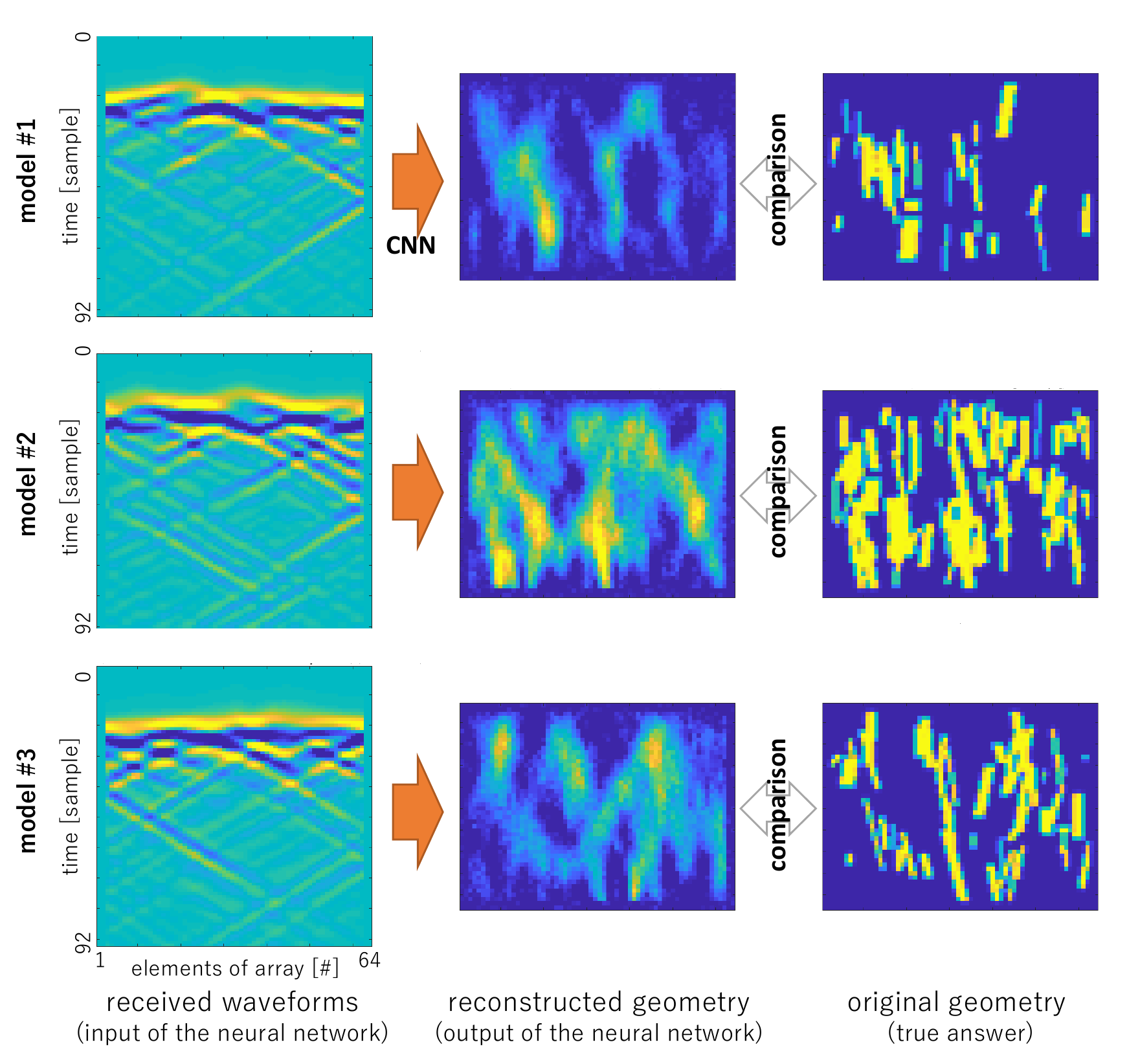}
    \captionsetup{margin=1.0cm,font={sc, footnotesize}}
    \caption{Result of simulation and output of CNN. We use (a) 2-D (spatial and time) received signal as the input of CNN and get the (b) 2D image as the output. We then compare the (b) output and (c) true answer.}
    \label{fig:fig3}
  \end{center}
\end{figure}

Figure \ref{fig:fig3} shows the estimation results. We used the received signals shown in Fig. \ref{fig:fig3} (a) as the test data. The results shown in Fig. \ref{fig:fig3} (b), the results with 64 elements array, show the good agreement of the true 2D arrangement shown in Fig. \ref{fig:fig3} (c). The proposed technique succeeded in estimating the geometry accurately and the results do not depend on the densities of the targets.

\begin{figure}[!h]
  \begin{center}
    \includegraphics[width=16.5cm]{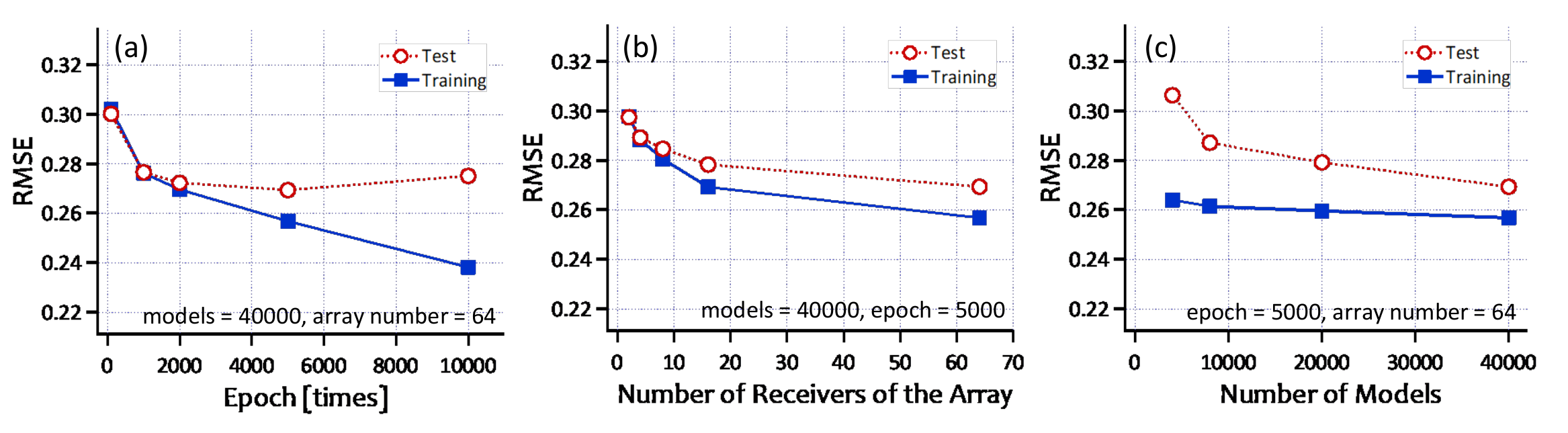}
    \captionsetup{margin=1.0cm,font={sc, footnotesize}}
    \caption{The dependence of (a) the number of epochs used in the CNN, (b) the number of receivers in the array, and (c) number of models used for the training and testing.}
    \label{fig:fig4}
  \end{center}
\end{figure}

Figure \ref{fig:fig4} shows the estimation results with several different parameters. We evaluated the result with root-mean-square error (RMSE). To realize the dependence of the parameters, we evaluate the proposed method with three different situations. Figure \ref{fig:fig4} (a) shows the dependence of the number of epochs used in the training. The larger number of epochs leads the better result with the training data (blue line shown in Fig. \ref{fig:fig4} (a)). However, we clearly see the over fitting when we use the test data without training data as shown in red line in Fig. \ref{fig:fig4} (a).

We also changed the number of receivers in the receiving array as shown in Fig. \ref{fig:fig4} (b). Because with the larger number of receivers, we can acquire the more spatial information, the larger number of receivers brings better estimation. 

We finally evaluated the dependence of the number of models used in the training as shown in Fig. \ref{fig:fig4} (c). The larger number of models brings better results. With the larger number of models, we succeeded in creating more accurate CNN. As shown in Fig. \ref{fig:fig3}, the proposed method shows the reasonable results.

In this study, we confirm the potential of the CNN for 2D ultrasound imaging with simple simulation parameters. As the future work, we will evaluate the dependence of density, sound speed and the effect of shear wave.


\section{Conclusions}

The potential of the 2D ultrasound imaging technique based on CNN was shown in this study. We considered the received signals at receiving array as a 2D image and retrieve the 2D geometry of the target medium of the propagation field by using CNN. As a result, the good agreements between the estimated images and true geometry were shown using simulation data. We believe that machine learning techniques such as the CNN can play an important role in the ultrasound imaging field. 


\section*{Acknowledgments}

The authors would like to thank Ryosuke O. Tachibana of The University of Tokyo for discussing the basic concept of preparing the data set for machine learning. This study was supported in part by KAKENHI (Grant Numbers 16K01431) from the Japan Society for the Promotion of Science (JSPS).


\end{document}